\documentclass[%
 reprint,
preprintnumbers,
 amsmath,amssymb,
 aps,
 prd,
]{revtex4-1}
\usepackage{graphicx}
\usepackage{dcolumn}

\usepackage[utf8]{inputenc} 

\newcommand{\be}{\begin{equation}}
\newcommand{\ee}{\end{equation}}
\newcommand{\ba}{\begin{eqnarray}}
\newcommand{\ea}{\end{eqnarray}}

\begin{document}

\preprint{INR-TH/2016-006}

\title{Dark matter component decaying after recombination: \\ 
Lensing constraints with Planck data}

\author{A. Chudaykin}
 \email{chudy@ms2.inr.ac.ru}
\affiliation{Institute for Nuclear Research of the Russian Academy of Sciences,
  Moscow 117312, Russia}%
\affiliation{Moscow Institute of Physics and Technology, 
  Dolgoprudny 141700, Russia}%

\author{D. Gorbunov}
 \email{gorby@ms2.inr.ac.ru}
\affiliation{Institute for Nuclear Research of the Russian Academy of Sciences,
  Moscow 117312, Russia}%
\affiliation{Moscow Institute of Physics and Technology, 
  Dolgoprudny 141700, Russia}%

\author{I. Tkachev}
 \email{tkachev@ms2.inr.ac.ru}
\affiliation{Institute for Nuclear Research of the Russian Academy of Sciences,
  Moscow 117312, Russia}%
\affiliation{Novosibirsk State University,
  Novosibirsk 630090, Russia}%

\begin{abstract}
It has been recently suggested~\cite{Berezhiani:2015yta} that emerging
tension between cosmological parameter values derived in high-redshift
(CMB anisotropy) and low-redshift (cluster counts, Hubble constant)
measurements can be reconciled in a model which contains subdominant
fraction of dark matter decaying after recombination.  We check
  the model against the CMB Planck data.  We find that lensing of the CMB
  anisotropies by the large-scale structure gives strong extra
  constraints on this model, limiting the fraction as $F<8\%$ at
  2\,$\sigma$ confidence level. However, investigating the combined data
  set of CMB and conflicting low-$z$ measurements, we obtain that 
  the model with $F\approx2\!-\!5$\% 
 exhibits better fit (by 1.5-3\,$\sigma$ depending on the lensing
 priors) compared to that of the concordance $\Lambda$CDM cosmological model.
\end{abstract}

\maketitle

\section{Introduction\label{sec:intro}}

The matter content of the Universe remains a mystery. Astronomical
observations and cosmological data resolve at least three components
in the matter sector: visible matter (baryons), dark matter (unknown
electrically neutral particles) and neutrinos~\footnote{Radiation and
  dark energy constitute another sectors.}. However, the dark matter
may be easily a multicomponent itself. Indeed, most models of the dark
matter and mechanisms of baryogenesis involve absolutely different
physics, so that order-of-magnitude equality between the contributions
of visible and dark matter to the present energy density of the
Universe is interpreted as a chance coincidence. Then why not to have
several different contributors to the dark matter sector?

To support this chain of reasoning, one can treat the neutrinos as a
second to baryon component in the "visible" or ordinary matter
sector. Both components are well recognizable in the cosmological data
analysis.  Similar situation may happen in the dark sector: there may
be several components as well. Moreover, they can be potentially
distinguishable. Furthermore, may be, the cosmological data already collected
allow us to observe hints of the dark matter component whose 
behavior differs from that of the canonical Cold Dark Matter
(CDM). Then, emerging discrepancies in fitting the $\Lambda$CDM
cosmological model to the growing stack of cosmological data might
signify the case.

Recent paper \cite{Berezhiani:2015yta} considering a subdominant dark
matter component decaying in the postrecombination epoch, is an
excellent example illustrating the above idea. In this model the total
dark matter amount gets reduced between the recombination and the
present epoch, which affects the cosmological observables. In fact,
the very model has been suggested to explain some emerging tensions
between the cosmological measurements ``at low redshifts'' and ``at
high redshifts''. Namely, it was argued in
Ref.\,\cite{Berezhiani:2015yta} that this model can explain low value
of the Hubble constant, $H_0=100\,h$\,km/s/Mpc, $h=0.6727\pm 0.0066$
\cite{Ade:2015xua} extracted from the analysis of the Cosmic Microwave
Background (CMB) anisotropy (Planck 2015, $\rm TT,TE,EE + lowP$ data) and
high values of the same constant extracted from the cosmic ruler
based on observations of astronomical Standard Candles,
$h=0.738\pm0.024$ \cite{Riess:2011yx}, $h=0.743\pm0.021$
\cite{Freedman:2012ny}.  Simultaneously, the model may explain a
tension between the cosmological constraints on $\sigma_8$ and
$\Omega_m$ from the CMB and from clusters as cosmological probes---the
cluster count data prefer lower values of these observables.


Measurements of these parameters can suffer from (unknown)
systematics, however, it must be unrelated for different observables
and experiments.  Hence the observation made in
Ref.\,\cite{Berezhiani:2015yta} may indeed be a hint of the
multicomponent dark matter and deserves further study.

The idea of a Decaying Dark Matter component has a long history,
starting apparently in the 1980s: see
e.g.\,\cite{Doroshkevich:1984,Flores:1986jn,Doroshkevich:1989bf}. There
are several varieties of it: CDM and a separate component decaying
into invisible radiation with lifetime shorter than the age of the
Universe, like in Ref.\,\cite{Berezhiani:2015yta}, a single unstable
CDM with lifetime exceeding the age of the Universe, like in 
Refs.\,\cite{Enqvist:2015ara,Blackadder:2015uta}, a two component DM,
where the heavier particles decay to lighter and invisible radiation,
like in Refs.\,\cite{Wang:2014ina,Cheng:2015dga,DelNobile:2015bqo}.   
This setup is argued to be possibly useful in not only explaining the
discrepancy between cosmological parameter estimates from 
CMB and matter clustering 
\cite{Aoyama:2014tga,Cheng:2015dga,Berezhiani:2015yta,Enqvist:2015ara},
Standard Candles\,\cite{Blackadder:2014wpa,Blackadder:2015uta}, 
but also, say, in relaxing the tension of CDM predictions with
observation of structures at (and of) small 
scales\,\cite{Aoyama:2014tga,Wang:2014ina,Cheng:2015dga} and in
understanding the origin of high-energy neutrino IceCube
events\,\cite{Anchordoqui:2015lqa}. The analysis performed in the
present paper for a particular variant of the 
model from Ref.~\cite{Berezhiani:2015yta} reveals an interesting
impact of decaying component on CMB, which is of general nature and
 is therefore expected to be relevant for other models as well.      

In Ref.~\cite{Berezhiani:2015yta} no real fitting to the Planck data
has been done actually.  Instead, to ensure that the model fits the
CMB anisotropy, the Planck derived values for all primary cosmological
parameters relevant at recombination were accepted and fixed. This
guarantees that anisotropies produced at the last scattering in both
models are identical. Furthermore, it was required that the angular
diameter distance to the last scattering should be the same for all
values of new parameters in the Decaying Dark Matter (DDM) model; 
namely, the sound horizon angle $100*\theta_s$ was fixed to the Planck
value as well. This guarantees that the observed CMB anisotropy
spectra in DDM model are {\it almost} identical to those in the
$\Lambda$CDM. The difference may appear only due to gravitational
distortions of spectra between last scattering and present. Though
such distortions are minor, they can be important with modern
cosmological data.  

There are two sources for these distortions. The first one is due to
integrated Sachs-Wolfe effect. It causes somewhat higher values of
anisotropy amplitudes $C_l$ at low multipoles $l$ in DDM as compared
to the Planck inspired $\Lambda$CDM. This is related to a larger
values of cosmological constant $\Lambda$ in the DDM assuming a flat
Universe. By itself this distortion is not very significant and was 
 effectively limited in~\cite{Berezhiani:2015yta} to levels
below cosmic variance by additional fitting to the supernova data. 

However, the second effect, the CMB distortion due to lensing by the
large scale structures, was not considered in
Ref.~\cite{Berezhiani:2015yta}. The difference in lensing power
between the  DDM
and  $\Lambda$CDM may be important since a part of the structure is
decaying in the former model, and may be observable with high-quality
data, such as Planck data. 

To fill this gap, in the present paper we fit DDM of
Ref.~\cite{Berezhiani:2015yta} to the complete Planck likelihood in
order to understand the importance of corresponding lensing
constraints. Our goal is to find out whether DDM may  indeed reconcile
cosmological measurements ``at low redshifts'' and ``at high
redshifts'' and whether it provides better description of the Universe
as compared to the $\Lambda$CDM model at the level of the current
data.  For this investigation we utilize the Planck 2015 CMB data
\cite{Ade:2015xua,Ade:2015len}, and the same constraints on the Hubble
constant~\cite{Riess:2011yx} and on $\sigma_8$ and $\Omega_m$ derived
from the Planck cluster counts~\cite{Ade:2015fva} which was used in
Ref.~\cite{Berezhiani:2015yta}.


\section{The model, data sets and procedure}
\label{sec:model_description}

\subsection{Decaying Dark Matter model}

Two component DDM model has two extra parameters, fraction of decaying
component in the total dark matter abundance, $F$, and its inverse
lifetime, or width, $\Gamma$.  To ensure a transparent transition to
the case of stable matter, the fraction $F$ is defined in terms of
{\it initial} energy densities $\omega_i \equiv \Omega_i h^2$ of
stable and decaying components in the following way $F \equiv{
\omega_{ddm}}/({\omega_{sdm} + \omega_{ddm}})$.   ``Initial'' here means
  the density which would have been measured if $\Gamma = 0$.
  Following Ref.~\cite{Berezhiani:2015yta} we also assume that the
  decay occurs into invisible massless particles (and does not produce
  too many photons) and normalize the width of the decaying component
  $\Gamma$ to km/s/Mpc, i.e. it is measured in the same units as
  $H_0$.

\subsection{Cosmological data sets}

To constrain this model we invariably employ TT,TE,EE Planck
likelihood for the power spectra at multipoles $l> 30$, as described
in~\cite{Ade:2015xua}. By itself it already contains the effects of
gravitational lensing of power spectra, which are most important for
us here.  Lensing reveals itself as smoothing of the acoustic
structure in power spectra, so the peaks become lower while the
troughs become higher.

We refer to the Planck measurements at low multipoles, $l < 30$, as
``lowP'' in notations of~\cite{Ade:2015xua}. This likelihood also
contains polarization data, which are crucial for us in what follows.

Since lensing is the main culprit for our investigation, we also
employ direct independent Planck measurements of the lensing power
spectrum $C_{l}^{\phi\phi}$. It is computed from the Planck's maps
using non-Gaussian (connected) parts of all 4-point correlation
functions (e.g. TTTT, TTEE, etc.)~\cite{Ade:2015len}. To avoid
confusion with the lensing extracted from TT,TE,EE we call 
the corresponding likelihood ``4lens''. This also highlights its origin in
4-point correlation functions. The Planck collaboration recommends
using this lensing likelihood since it ``constrains the lensing amplitude
more directly and more tightly''~\cite{Ade:2015len}.

For the low-redshift data sets, which are currently conflicting with
the base Planck $\Lambda$CDM cosmology, we use the same data sets as in
Ref.~\cite{Berezhiani:2015yta}. Namely, we use a direct astrophysical
measurement of the Hubble constant by Riess {\it et
  al.}~\cite{Riess:2011yx}, and indicate it as ${\rm H_0}$ in the
descriptions of data combinations.  As for the data on the galaxy
cluster counts, we adopt and add Planck results~\cite{Ade:2015fva}. We
refer to this Planck data set as ``CL.''

\begin{table}
\begin{center}
{\renewcommand{\arraystretch}{1.2}%
\begin{tabular}{|c| l |}
\hline 
Tag & ~~~~~~~~  ~~~~~~~~ {Data set} \\
\hline 
Pol & ~$\rm (TT,TE,EE + H_0 + CL) + lowP$\\
Lens & ~$\rm (TT,TE,EE + H_0 + CL) + 4lens $\\
Pol + Lens &  ~$\rm (TT,TE,EE + H_0 + CL) + lowP+4lens$~\\
\hline 
\end{tabular}
}
\end{center}
\caption{
Data sets used in our analysis and their tags.
\label{tab:sets}
}
\end{table}

In our discussion of the main results, we always employ the following block of
data $\rm (TT,TE,EE + H_0 + CL)$. However, lensing amplitude which is
contained here is conflicting with the one from 4lens,
see~\cite{Ade:2015xua,Ade:2015len}. Therefore, we consider this block
in three different combinations with lowP, 4lens and lowP+4lens, as
summarized in Table~\ref{tab:sets}. Tags for these combinations
reflect main relevant differences between them.


\subsection{Numerical procedure}

All relevant cosmological calculations have been carried out
numerically using the CLASS Boltzmann
code~\cite{Lesgourgues:2011re,Blas:2011rf}. The parameter space is
explored using the Markov Chain Monte-Carlo technique with the Monte
Python package~\cite{Audren:2012wb}.  The two-component DDM model is
predefined in the both numerical tools.  Eight primary cosmological
parameters have been varied. Out of eight, two are specific for the
DDM model: the fraction $F$ and the width $\Gamma$. The remaining six
parameters are standard: the angular size of the sound horizon $r_s$
at last-scattering $\theta_*\equiv 100\times r_s(z_*)/D_A(z_*)$, the
baryon density $\omega_b $, initial CDM density $\omega_{cdm} =
\omega_{sdm} + \omega_{ddm}$, the optical depth $\tau$, the squared
amplitude $A_s$ and tilt $n_s$ of the power spectrum of primordial
scalar perturbations. In the numerical codes we run the perturbations 
 at the linear regime only. We have checked (by switching on the
 corresponding option in CLASS code) that our main
  results change very mildly with account of the nonlinear
  corrections to the lensing potentials, which happens to be most
    important in testing the model with CMB data, as we show below. 
Nonlinear contributions to the matter power spectrum $P(k)$, calculated
with CLASS, are below 1\% at the scales relevant for the estimates of 
parameter $\sigma_8$.    
We are planning detailed study of these and other delicate effects
associated with the nonlinear evolution in a forthcoming paper.  
  In this study we take the Universe to be spatially
flat, neglect the possible tensor perturbations and put the sum of
active neutrino masses equal to $\sum m_\nu=0.06$\,eV assuming 
a nondegenerate normal hierarchy pattern.


\section{Constraints on DDM}
\label{sec:fits}

\subsection{Planck data only}

\begin{figure} [!htb]
\centerline{
\includegraphics[keepaspectratio,width=\columnwidth]{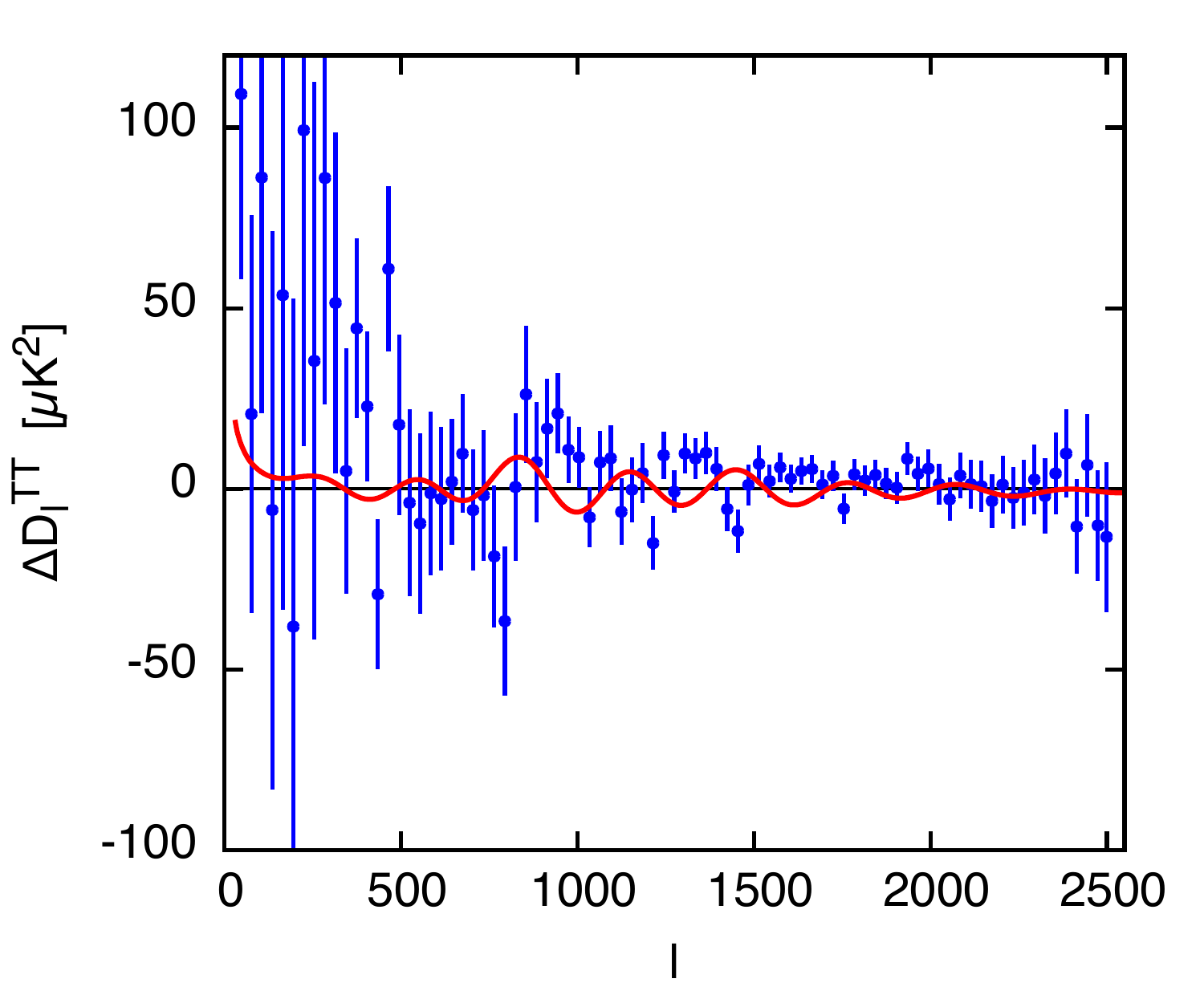}
}
\caption{\label{fig:44} Data points with error bars show residuals
  after subtraction from the measured TT power spectrum the
  $\Lambda$CDM model prediction with the best-fit parameters from
  $\rm TT,TE,EE + lowP$ analysis.  Solid curve corresponds to the
  difference between TT spectra in DDM ($F=0.1$,
  $\Gamma=2000$\,km/s/Mpc) and the same $\Lambda$CDM model.}
\end{figure}

First, we would like to  visualize the important role of lensing
effects in TT power spectra. For this we plot the difference between
predictions of DDM 
($F=0.1$, $\Gamma=2000$\,km/s/Mpc) and of $\Lambda$CDM models, while
taking other parameters being fixed by the best fit 
to the $\rm TT,TE,EE + lowP$ data set. This difference 
is shown in Fig.~\ref{fig:44} 
by the solid
curve. The best-fit $\Lambda$CDM model  spectrum
is also subtracted from the
data, residuals are shown by dots with error bars. 
We see that the difference between DDM and CDM is appreciable at the
level of presently achieved precision and, therefore, lensing should be
included in constraining the DDM models. We also see that 
while generically the lensing is accounted for in both models 
(i.e. residuals are small), the agreement with data is not perfect even for the
base $\Lambda$CDM model: the data points after subtraction oscillate
coherently  in the vicinity of zero. 
Somewhat more lensing power is required to fit
the data as compared to the theoretical prediction in the $\Lambda$CDM
model, the disagreement is at $2\sigma$ level~\cite{Ade:2015len}.
While the amplitude of the difference between the 
DDM and $\Lambda$CDM (solid curve)
is comparable to the deviations of residuals, it is out of phase. 
This reflects even weaker lensing power in DDM since the
large-scale structure is decaying at late times. As a result, fitting
to TT,TE,EE Planck likelihood alone restricts DDM to the range $F <
0.07$ at $2\sigma$ level.  This is the key observation missed in
  Ref.\,\cite{Berezhiani:2015yta}.

By itself, lowP likelihood does not gives strong constraints. However,
the situation is little bit more tricky and its role is important
together with TT,TE,EE. It works as follows. The lack of lensing power
in theoretical predictions to match TT data pushes the fit to 
a higher amplitude of primordial spectra, $A_s$. To
compensate for a simultaneously growing amplitude of $C_l^{TT}$, this
in turn requires larger optical depth, $\tau$, but the latter is
limited by polarization data in the lowP likelihood. As a result, DDM
is even more limited in the $\rm TT,TE,EE + lowP$ likelihood, 
{$F <0.04$ at $2\sigma$ level.} 

\begin{figure}[!htb]
\includegraphics[width=\columnwidth]{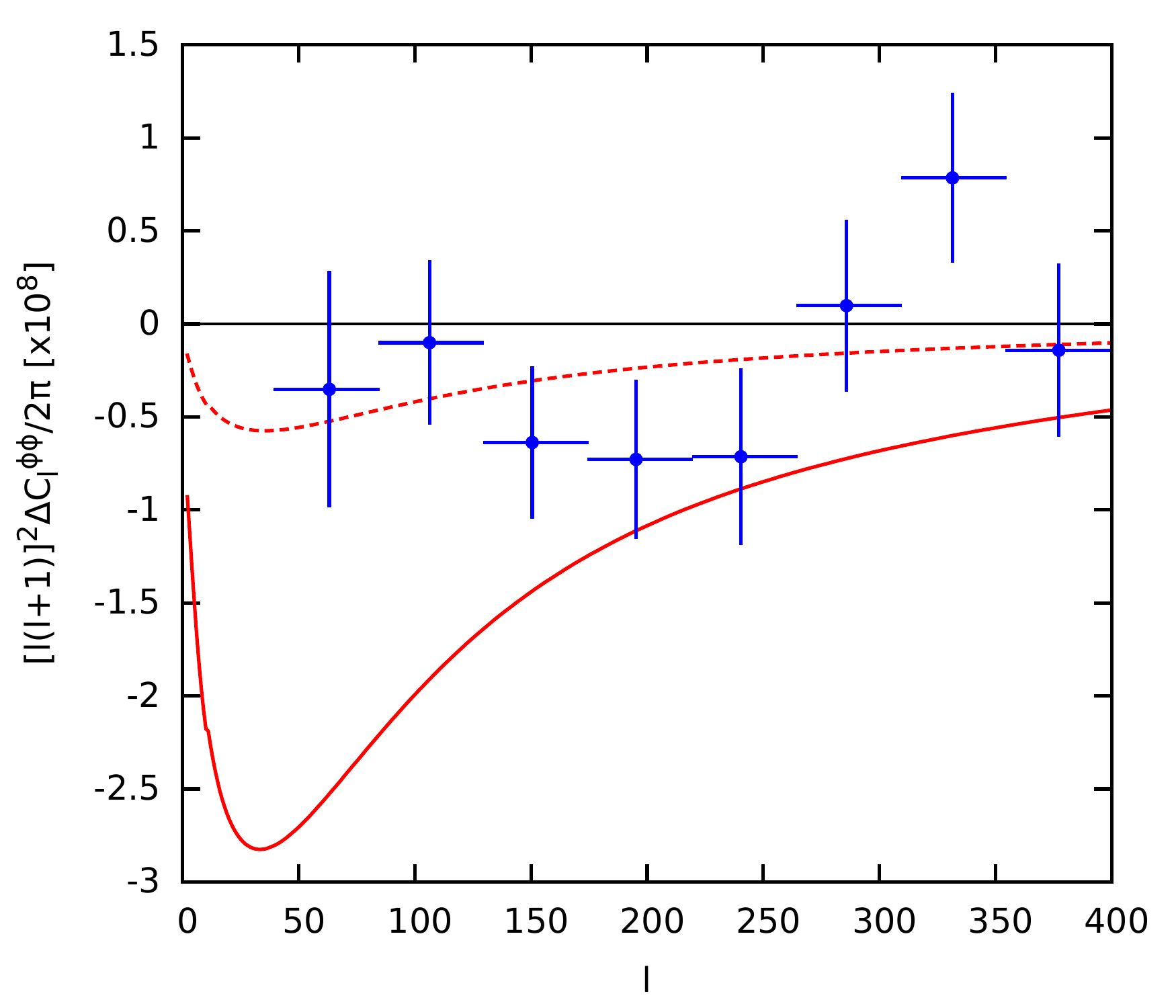}
\caption{
\label{bis}
Data points with error bars show residuals after subtraction from the
directly measured lensing power spectrum the $\Lambda$CDM model
prediction with the best-fit parameters from $\rm TT,TE,EE + lowP$
analysis. Solid and dotted curves show the difference between
$C_{l}^{\phi\phi}$ in DDM and the same $\Lambda$CDM model: for the
solid curve $F=0.1$, $\Gamma=2000$\,km/s/Mpc were used as the DDM
parameters, while the dotted curve corresponds to the best-fit DDM in
the $\rm TT,TE,EE + lowP + 4lens$ analysis.  }
\end{figure}

The situation with the lensing power spectrum $C_{l}^{\phi\phi}$ is
 opposite. Theoretical predictions based on the $\Lambda$CDM model
 also disagree with the Planck data here, but now less power is needed
 to explain data~\cite{Ade:2015xua,Ade:2015len}, i.e.  this direct
 lensing power spectrum slightly favors DDM  as compared to
   $\Lambda$CDM, see Fig.~\ref{bis}. As a result, upper boundary for $F$ gets little bit
 more relaxed, { $F <0.08$ at $2\sigma$ level}.

\subsection{Planck data and conflicting low-$z$ measurements}

Now we combine Planck data with conflicting low redshift measurements
of $H_0$, $\Omega_m$ and $\sigma_8$. Since DDM is restricted by
lensing, and the lensing acts in the opposite directions in TT,TE,EE
and $C_{l}^{\phi\phi}$ likelihoods, we scrutinize the model using
three data combinations listed in Table~\ref{tab:sets}.

\begin{figure}[!htb]
\includegraphics[keepaspectratio,width=\columnwidth]{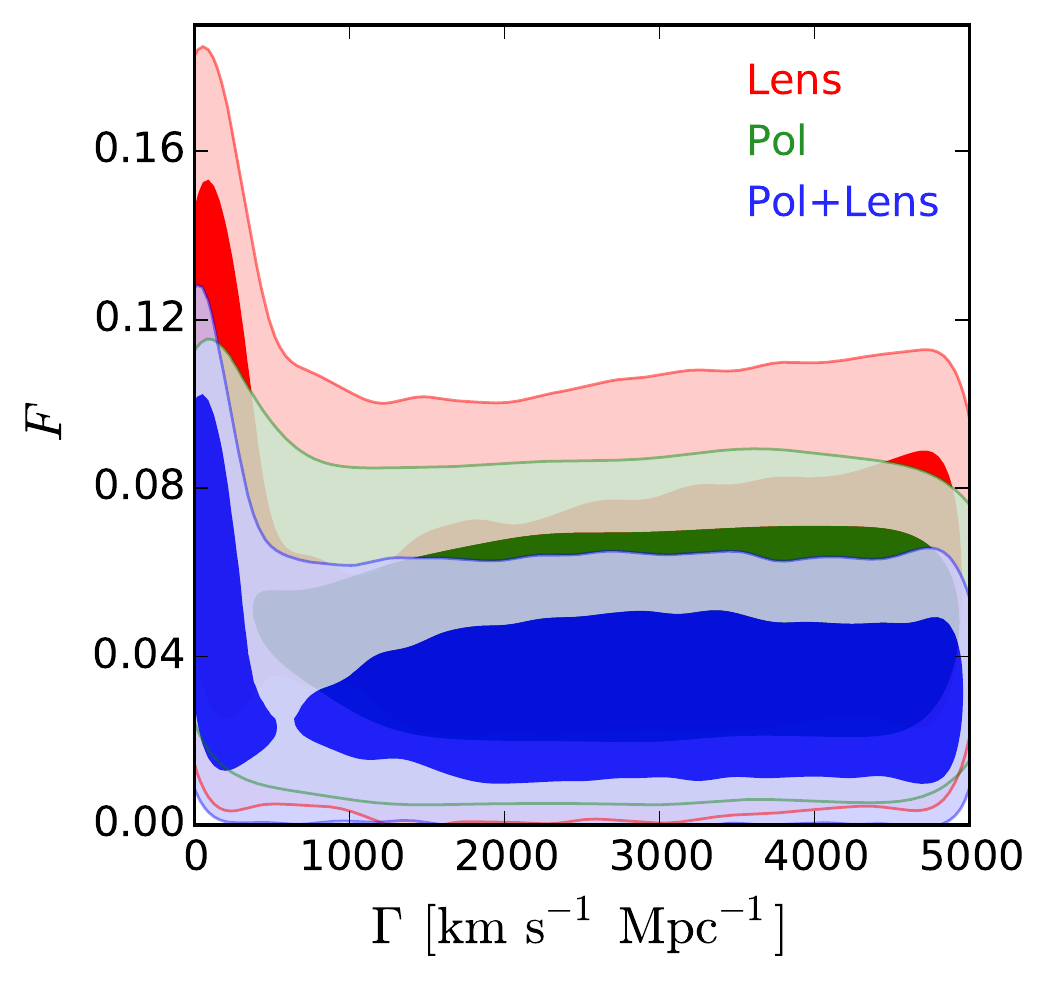}
\caption{\label{fig:7} 
Posterior distributions ($1\,\sigma$ and $2\,\sigma$ contours) of parameters $F$,  $\Gamma$ in DDM model. Tags are described in Table~\ref{tab:sets}.}
\end{figure}

\begin{figure}[!htb]
\includegraphics[keepaspectratio,width=\columnwidth]{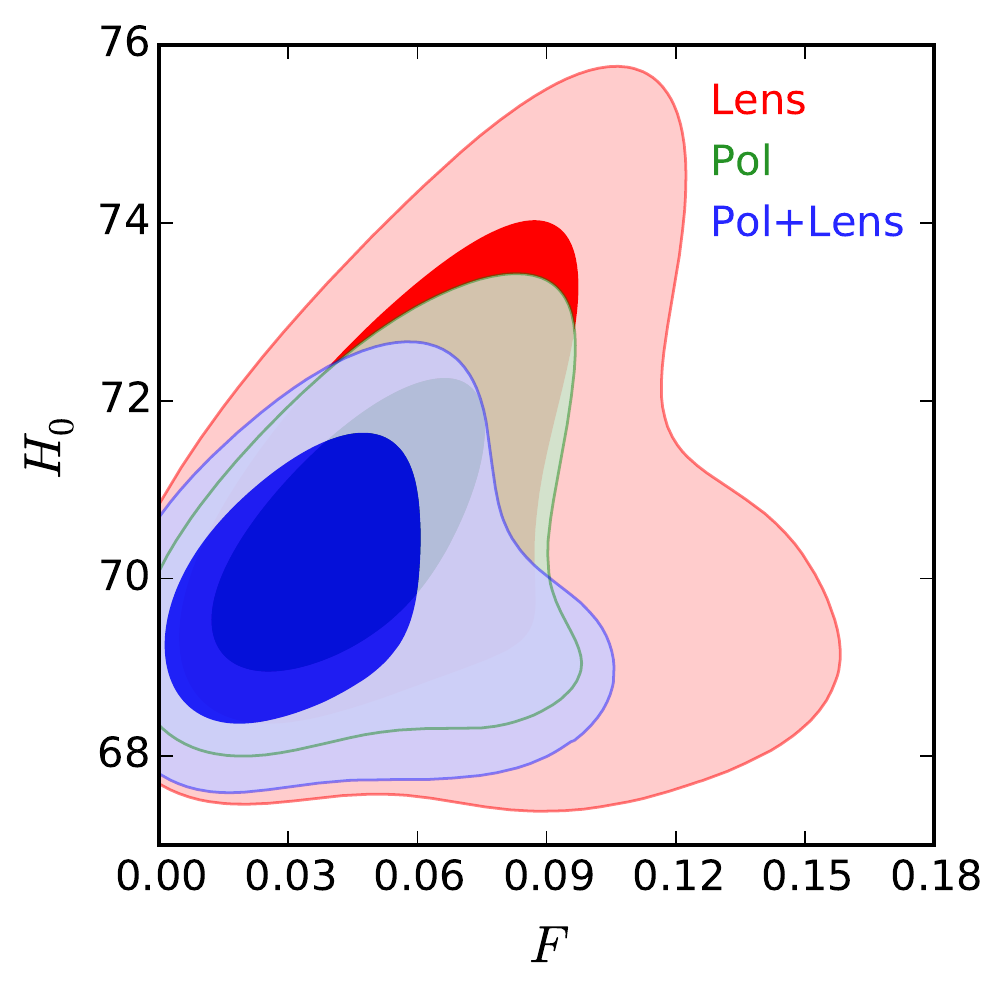}
\caption{\label{fig:8} Same as Fig.~\ref{fig:7} but for $H_{0}$ and
  $F$.}
\end{figure}

\begin{figure}[!htb]
\includegraphics[keepaspectratio,width=\columnwidth]{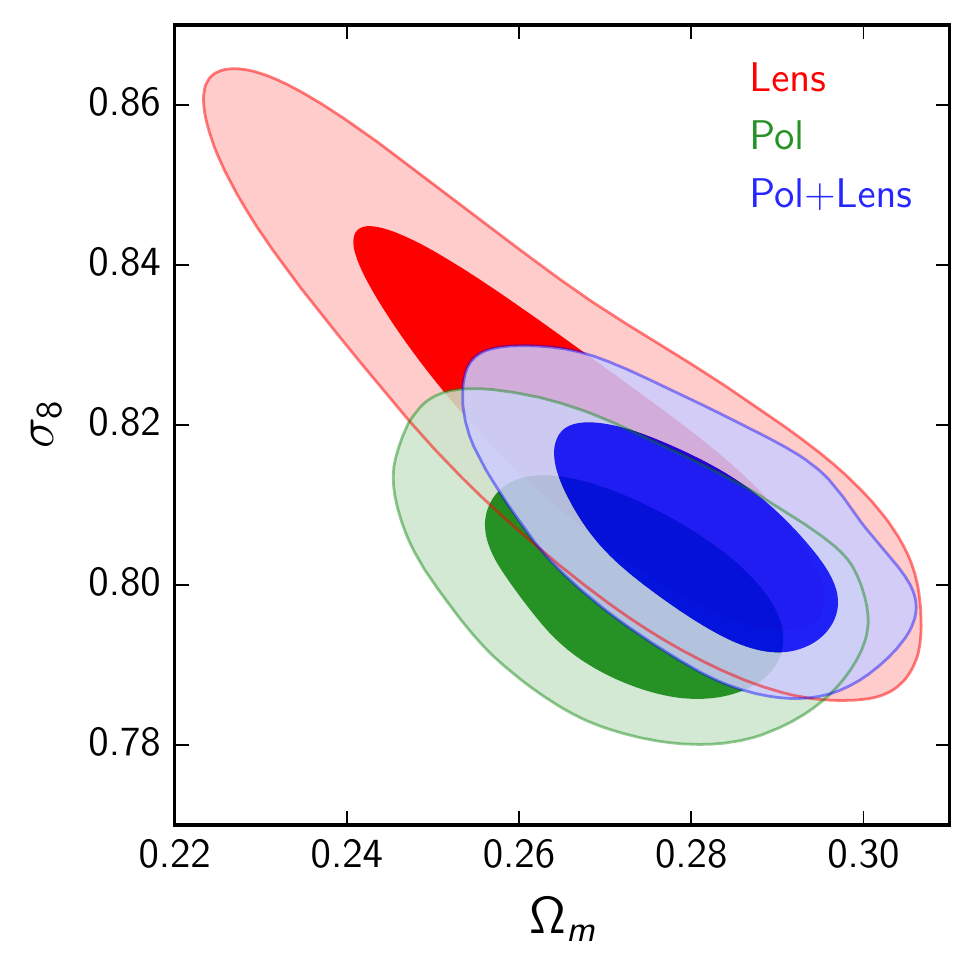}
\caption{\label{fig:9} 
Same as Fig.~\ref{fig:7}  but for $\sigma_8$ and $\Omega_m$. 
}
\end{figure}

Corresponding constraints on DDM parameters and on conflicting
cosmological parameters $H_0$, $\Omega_m$ and $\sigma_8$ are presented
in Figs.~\ref{fig:7} - \ref{fig:9}. 
One can see in Fig.\,\ref{fig:7} that preferred fraction of DDM
$F\simeq 2-5\%$ is almost independent of $\Gamma$ in the range
$1000\,\text{km/(s\,Mpc)}\lesssim \Gamma\lesssim
5000\,\text{km/(s\,Mpc)}$ with DDM completely disappearing between
recombination and present epoch, $\Gamma\gg H_0$. In this range of
$\Gamma$ the models with minuscule fraction of decaying component,
$F=0$, are within the 2\,$\sigma$ contours for the lens and for
combined data sets, but outside for the Pol.  The increase in allowed
$F$ at smaller values of $\Gamma$, where the DDM lifetime grows,
corresponds to the situation, where a part of the DDM particles survive 
in the late Universe. In particular, at $\Gamma\ll H_0$ the DDM is
indistinguishable from the stable DM. The region of small $\Gamma$ is
not resolved in our figures and deserves special study beyond the
scope of this paper.  In Fig.\,\ref{fig:8}, the allowed at 2\,$\sigma$
regions with highest values of $F$ and smallest values of $H_0$ map in
Fig.\,\ref{fig:7} to the regions with longer lifetimes, $\Gamma\ll
1000\,\text{km/(s\,Mpc)}$.  In Fig.\,\ref{fig:9} the regions of
highest allowed values of $\sigma_8$ and lowest values of $\Omega_m$
map in Fig.\,\ref{fig:7} to the region of $1000\,\text{km/(s\,Mpc)}
\lesssim \Gamma\lesssim 5000\,\text{km/(s\,Mpc)}$ and the highest
allowed values of $F$.

We also observe that the proper fit to CMB data reveals
a significantly smaller fraction of DDM as compared to the results of
Ref.\,\cite{Berezhiani:2015yta}, and  the favored  model
parameter values in Fig.\,\ref{fig:7} are well outside the
$2\,\sigma$-region presented in Ref.\,\cite{Berezhiani:2015yta}. 
However, our fits still indicate nonzero  $F$, and absence of the decaying
component is disfavored.

\begin{table}
\begin{center}
{\renewcommand{\arraystretch}{1.2}%
\begin{tabular}{|c|c|c|c|}
\hline
Data set & ~$\Delta \chi^2$~ & ~P value~ & Improvement \\
\hline 
Pol & 3.84 & 0.1466 & $1.45\sigma$ \\
Lens & 11.68  & 0.0029 & $2.89\sigma$ \\
Pol + Lens & 8.74 & 0.0126 & $2.47\sigma$ \\
\hline 
\end{tabular}
}
\end{center}
\caption{
Improvement of DDM over $\Lambda$CDM in three data sets considered taking into account 2 extra degrees of freedom in DDM.
\label{tab:1}
}
\end{table}

To understand quantitatively which model ($\Lambda$CDM or DDM) is
preferable according to the 
cosmological data, we compare the differences in logarithmic likelihoods
$\log L$ calculated for these two models in their respective best-fit
points for the same data sets. Each difference $2\cdot\Delta\log L$ is
distributed as $\chi^{2}$ with an effective number of degrees of
freedom equal to the difference in the number of fitting parameters in
these two models, which is 2, corresponding to two extra parameters
$F$ and $\Gamma$ in DDM. Resulting improvements of the DDM over the 
$\Lambda$CDM are displayed in Table~\ref{tab:1}.

There is an improvement of the DDM over 
$\Lambda$CDM, and hence the DDM indeed describes the
cosmological data better, as suggested in Ref.\,\cite{Berezhiani:2015yta}. 
However, the improvement is not very significant 
because of the key disagreement between theory and CMB measurements
displayed in Fig.~\ref{fig:44}. It enters all our data sets, and is
worse for the DDM than for the $\Lambda$CDM.

In principle, the corresponding constraints,  when strengthened, 
can rule out DDM, but currently they may be judged as
rather harmless. Indeed, it has to be understood first why
$\Lambda$CDM is also in tension with the lensing here, and then the 
disagreement has to be resolved. Before that, it is unreasonable to
make strong conclusions in either direction.

\section{Conclusions}
\label{sec:conclusions}

We confirm that DDM provides a better description of the CMB and
low-$z$ measurements of cosmological parameters. However, the fraction
of the DDM is much smaller than previously claimed\,\cite{Berezhiani:2015yta}.  

Namely, we have found 
that lensing of CMB anisotropies by large-scale structures, contained
in the Planck data, strongly constrains the two-component DDM model. 
Essentially, lensing is measured twice. First, as smoothing of the
acoustic peaks in TT power spectra and, second, directly in the
lensing power spectrum $C_{l}^{\phi\phi}$. Both measurements are
slightly conflicting with predictions of the $\Lambda$CDM model. And,
in a sense, conflict between them can be also considered as conflict
between low- and high-$z$ measurements. $\Lambda$CDM predicts smaller
lensing power as compared to what is required by TT+lowP power
spectrum results and larger lensing power compared to
$C_{l}^{\phi\phi}$ results. Therefore, the former likelihood strongly
restricts DDM, while the latter actually favors.  This conclusion
is rather generic and CMB measurements themselves must constrain 
other models with DDM component, see Introduction, as well.

Then, with this conflict
in the backyard, we have analyzed whether DDM is able to reconcile
the Planck-inspired $H_0$, $\Omega_m$, and $\sigma_8$ values with their
conflicting low-redshift measurements. Improvement of DDM over
$\Lambda$CDM is observed, but it is not very significant 
with the current data. We feel 
that for the final verdict it is highly important to understand the
source of the "lensing conflict" in the Planck data.

\vskip 0.3cm 
We thank Z.\,Berezhiani, A.\,Dolgov and J.\,Lesgourgues for valuable
correspondence.  
The work of A.Ch. and D.G. 
has been supported by the Russian Foundation for Basic Research grant
14-02-00894. All computations in the work were made with the MVS-10P
  supercomputer of the 
Joint Supercomputer Center of the Russian Academy of Sciences (JSCC
RAS). 
\bibliography{refs_r} 

\end{document}